\documentclass[aps,prl,twocolumn,superscriptaddress]{revtex4-2}

\usepackage[utf8]{inputenc}
\usepackage{bbm}

\usepackage{placeins}
\usepackage{xcolor}
\usepackage{graphicx}
\usepackage{hyperref}
\usepackage[export]{adjustbox}

\usepackage{mathtools}
\usepackage{amsmath}
\usepackage{amssymb}
\usepackage{physics}
\usepackage{braket}

\hypersetup{pdfborder = {0 0 0}}
\hypersetup{colorlinks, citecolor = blue, filecolor = blue, linkcolor = blue, urlcolor = blue}

\begin{document}
\title{
Two-particle tunneling and the impact of interaction}

\author{Jonathan Brugger}
\affiliation{Physikalisches Institut, Albert-Ludwigs-Universität Freiburg,\\ Hermann-Herder-Straße 3, 79104 Freiburg, Germany}
\affiliation{EUCOR Centre for Quantum Science and Quantum Computing, Albert-Ludwigs-Universität Freiburg,\\ Hermann-Herder-Straße 3, 79104 Freiburg, Germany}

\author{Christoph Dittel}
\affiliation{Physikalisches Institut, Albert-Ludwigs-Universität Freiburg,\\ Hermann-Herder-Straße 3, 79104 Freiburg, Germany}
\affiliation{EUCOR Centre for Quantum Science and Quantum Computing, Albert-Ludwigs-Universität Freiburg,\\ Hermann-Herder-Straße 3, 79104 Freiburg, Germany}
\affiliation{Freiburg Institute for Advanced Studies, Albert-Ludwigs-Universität Freiburg,\\ Albertstraße 19, 79104 Freiburg, Germany}

\author{Andreas Buchleitner}
\email{abu@uni-freiburg.de}
\affiliation{Physikalisches Institut, Albert-Ludwigs-Universität Freiburg,\\ Hermann-Herder-Straße 3, 79104 Freiburg, Germany}
\affiliation{EUCOR Centre for Quantum Science and Quantum Computing, Albert-Ludwigs-Universität Freiburg,\\ Hermann-Herder-Straße 3, 79104 Freiburg, Germany}

\begin{abstract}
We analyze the tunneling of two 
bosons in a double-well, for 
contact, soft-, and hard-core Coulomb 
interaction of tunable strength. 
Transitions from correlated to uncorrelated tunneling 
of the 
left well's 
two-particle 
ground state are 
due to resonances with 
states with one particle in either well. Their abundance and dependence on the interaction strength
is indicative of the interaction type.
\end{abstract}

\date{\today}

\maketitle

Tunnelling processes \cite{Hund1927}, which mediate transitions between classically separate regions of position \cite{CohenTannoudjiQM1}, momentum or phase space \cite{Davis1981}, 
are a paradigmatic property 
of the quantum realm, and give rise to a multitude of spectral features \cite{abu2002}, dynamical phenomena and technological applications across all domains of physics, 
from nuclear decay \cite{Gamov1928} to surface tunneling microscopy \cite{STM}. In transport theory \cite{benjacob1985,QuantumTransport1,QuantumTransport2} as well as in 
atomic and molecular fragmentation 
dynamics \cite{Walker1993,PhysRevLett.76.2654}, 
where, in general, multiple 
particles are involved, tunneling and interaction
processes coexist, and their mutual interplay  \cite{abu2004} depends on both, the potential landscape and the particles' interaction potential. While this interplay can have crucial impact
on the speed \cite{Dengis2024} and/or on the level of synchronicity \cite{Winkler2006,Murmann2015,FloquetTunneling,NOON2021,NOON2022} of the particles' tunneling transitions and hence, e.g., on molecular fragmentation 
or reconformation processes \cite{QuantumTunnelingChemistry}, 
we lack a systematic understanding of how different interaction potentials -- short vs. long range, soft- vs. hard-core -- imprint the tunneling dynamics.

It is the purpose of our present 
contribution to help develop a general understanding of the similarities and differences 
of the tunneling dynamics as induced by 
distinct interaction types. This becomes feasible through 
a versatile finite element \cite{BartelsFEM, SunZhouFEM}
representation of the underlying eigenvalue problem, which can account for variable interaction potentials and allows to resolve the essential spectral structure
of the problem, with high numerical accuracy \cite{DissertationJB}. We will see that, in the most elementary and rather ubiquitous scenario of a one-dimensional double-well 
potential \cite{CohenTannoudjiQM1,Davis1981,abu2002,AsymDWPotential,StefanHunnPRA,StefanHunnDiss,Murmann2015,Brugger2023},
two interacting bosons launched in an isolated well's two-particle ground state exhibit a transition from correlated to uncorrelated tunneling whenever the interaction shifts 
the initial state into resonance with an eigenstate where one particle resides in either well. While this condition is the same irrespective of the specific form of the interaction 
potential, it is met more frequently for long-range interactions, and at smaller interaction strengths for the hard-core Coulomb potential, as we will elaborate upon in more detail below.

\begin{figure}[t!]
	\centering
	\includegraphics[width=\linewidth]{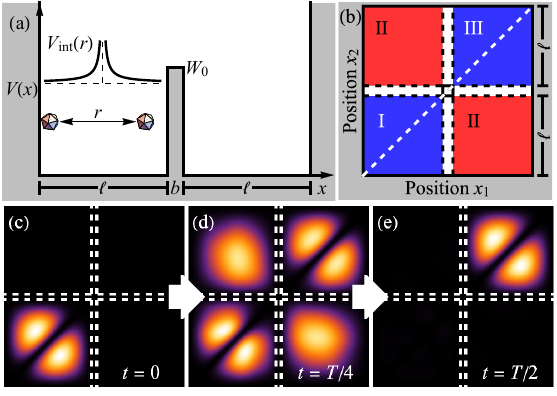}
	\caption{Configuration space and two-particle densities of two interacting bosons confined  (a) to two 
	symmetric 
	wells of individual width $\ell$, separated by a barrier of height $W_0$ and width $b$. 
	(b) Two-particle configuration space associated with (a), spanned by the single-particle coordinates $x_{1,2}$.
	Regions I and III (region II) represent 
	particles occupying the same (different) well(s). The dashed white line indicates $x_1 = x_2$. (c) 
	Probability density of the ground state $\ket{g_\mathrm{L}}$ of two interacting bosons in the isolated left well (i.e., $W_0 = \infty$), for repulsive hard-core Coulomb interaction $V_{\rm int}$
	as sketched in (a). (d--e) Time-evolved densities at $t=T/4, T/2$, $T$ the tunneling 
	period, with initial condition 
	(c), after a potential quench to finite $W_0$ at $t=0$.
	}\label{Fig1}
\end{figure}

Consider two interacting, indistinguishable bosons in a symmetric, one-dimensional double-well potential $V(x)$  \cite{AsymDWPotential,StefanHunnDiss,StefanHunnPRA,Brugger2023}, 
consisting of two flat wells of 
length $\ell$, separated by a barrier of height $W_0$ and width $b$ [see Fig.~\ref{Fig1}(a)]. 
In
natural units \cite{StefanHunnPRA} ($\hbar \equiv 1$ and $m \equiv 1/2$), the single-particle Hamiltonian reads $H_\mathrm{1P}(x) = - \partial^2/\partial x^2 + V(x)$, and 
the two-particle Hamiltonian is given by
\begin{gather}
	H_\mathrm{2P}(\vec{x}) = H_\mathrm{1P}(x_1) + H_\mathrm{1P}(x_2)  + U \, V_\mathrm{int}(x_1 - x_2),\label{Eq1}
\end{gather}
with $\vec{x}=(x_1,x_2)$, $V_\mathrm{int}(r)$ the interaction potential, and $U$ the interaction strength. If all distances are measured in units of an experimentally given 
length scale $\mathcal{L}$, then $H_\mathrm{2P}(\vec{x})$ and all other energies -- such as the barrier height $W_0$ -- are given in units of $\mathcal{L}^{-2}$, while time
is given in units of $\mathcal{L}^2$, and the interaction strength and potential in units of $\mathcal{L}^{-1}$.

\begin{figure}[t!]
	\centering
	\includegraphics[width=\linewidth]{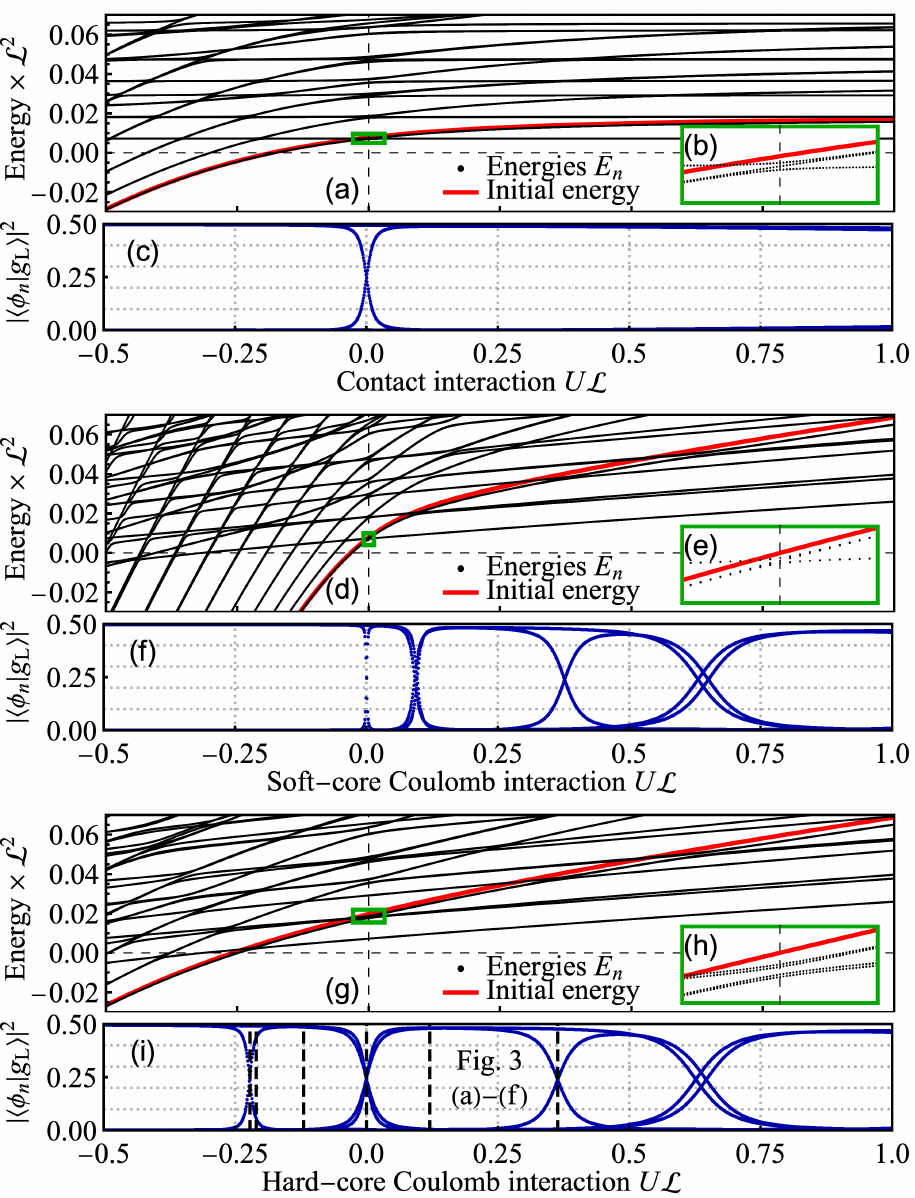}
	\caption{(a,d,g) Parametric evolution of the energy levels $E_n$ (black) of two interacting bosons in the double-well potential depicted in Fig.~\ref{Fig1}, 
	as a function of the scaled interaction strength $U\mathcal{L}$, for
	contact, soft-, and hard-core Coulomb interaction (from top to bottom), respectively.
	The energy of the two-particle ground state $\ket{g_\mathrm{L}}$ of the isolated left well is represented by the red line.
	The insets (b,e,h) zoom into the avoided crossings at 
	$U=0$ (green boxes). 
	(c,f,i) illustrate the parametric evolution of the decomposition of $\ket{g_\mathrm{L}}$ into the dominant eigenstates $\ket{\phi_n}$ associated with the $E_n$,
	by their respective weigths $\left|\braket{\phi_n|g_\mathrm{L}}\right|^2$. 
	Dashed vertical lines in (i) (from left to right) identify the scaled interaction strengths giving rise to the dynamics depicted in Fig.~\ref{Fig3} (a-f).
	}
	\label{Fig2}
\end{figure}

\begin{figure}[t!]
	\centering
	\includegraphics[width=\linewidth]{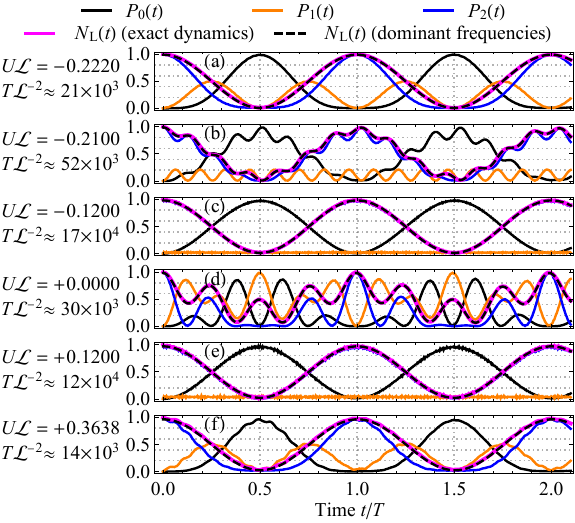}
	\caption{Time evolution of the probabilities $P_n(t)$ to find $n$ particles, and of the expected particle number $N_\mathrm{L}(t)$,
	in the left well (see legend), for hard-core Coulomb interaction of increasing strength $U$, from (a) (attractive) to (f) (repulsive) [the considered $U$ values are also indicated by vertical dashed lines in 
	Figs.~\ref{Fig2}(i) and \ref{Fig4}(e,f)]. 
	In (c,e), the particles exhibit 
	correlated 
	tunneling, with $P_1(t) \approx 0$ for all times, 
	while in (a,b,d,f) they tunnel individually -- 
	at those values of $U$ 
	where the initial two-particle state moves into resonance [avoided crossings in Fig.~\ref{Fig2}(g)] with an eigenstate with one particle in either well of the potential. Note that, in 
	either situation [e.g., (a,c)], the behaviour of $N_\mathrm{L}$ can be qualitatively similar, while 
	$P_n(t)$ reflects the level of correlation.
	In all cases, 
	$N_\mathrm{L}(t)$ is
	well approximated (black dashed line)
	by \eqref{Eq5}, using the dominant frequencies $\omega_k$ with amplitudes $A_k \geq 0.01$ (see Fig.~\ref{Fig4}). 
	Note the different scaled tunneling times $T\mathcal{L}^{-2}$, determined by the 
	smallest dominant frequency, 
	given on the left. Qualitatively similar results are observed 
	for contact and soft-core Coulomb interaction. 
	}\label{Fig3}
\end{figure}

\begin{figure}[t!]
	\centering
	\includegraphics[width=\linewidth]{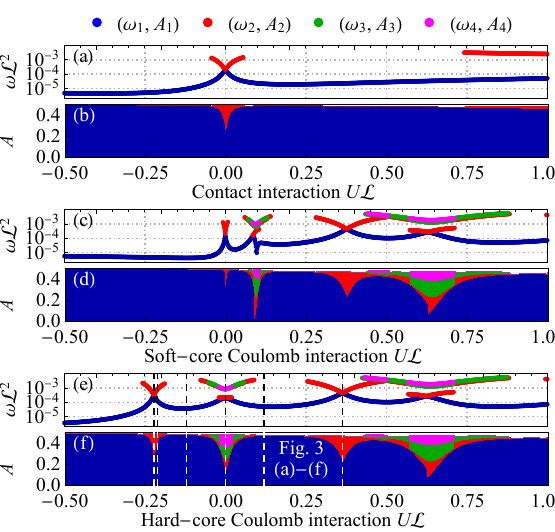}
	\caption{Dominant frequencies $\omega_k$ and amplitudes $A_k$ in the time evolution \eqref{Eq5} of the normalized particle number $N_\mathrm{L}(t)$ 
	in the left well, as a function of the scaled interaction strength $U\mathcal{L}$, for (a,b) contact, (c,d) soft-,
	and (e,f) hard-core Coulomb interaction. (a,c,e) show all dominant frequencies $\omega_k$ (in different colors), 
	(b,d,f) 
	stacked bar charts of the associated 
	amplitudes $A_k$ (with $A_k \geq 0.01$; same color coding). Note that, 
	since $\sum_k A_k = 1/2$, the color coded
	amplitudes add up to approx. $1/2$, for any value of $U\mathcal{L}$. The vertical dashed lines (from left to right) in (e,f) identify those $U\mathcal{L}$ giving rise to the 
	dynamics depicted in Figs.~\ref{Fig3} (a-f).
	}\label{Fig4}
\end{figure}

The 
parameters $\ell$, $W_0$, and $b$ determine the 
energy level structure within either well, as well as 
the coupling between them
and, consequently, the tunneling 
time scale. We 
fix $l = 50 \, \mathcal{L}$, $b = 3 \, \mathcal{L}$,  $W_0 = 0.3 \, \mathcal{L}^{-2}$. 
Given this potential landscape, we consider three different interaction potentials: a short-range contact interaction $V_\mathrm{int} (r) = \delta(r)$ \cite{Bethe, DifferentInteractionsContactAnalytical, DifferentInteractionsApproximations}, 
and a hard-core, 
$V_\mathrm{int} (r) = 1  / |r|$ \cite{BuchleitnerEtAl2}, as well as 
a soft-core, $V_\mathrm{int} (r) = 1  / \sqrt{r^2+\Delta^2}$ \cite{SoftCoulomb1981, SoftCoulomb1985, 
DifferentInteractionsApproximations}, with $\Delta = 1 \, \mathcal{L}$, long-range Coulomb interaction.
The 
scaled interaction strength is chosen in the range $U \mathcal{L} \in [-0.5, 1]$, which is
experimentally realized 
either by direct
variation of 
$U$, as done hereafter,
or of
the length scale $\mathcal{L}$ [and of the sign of $U$; positive (negative) values correspond to repulsive (attractive) interaction]  \footnote{{Alternatively, also the potential wells' width $\ell$ can 
be changed, while keeping $b$, $W_0$, and $U$ constant. This yields similar results to those in Figs.~\ref{Fig2}-\ref{Fig4} \cite{DissertationJB}.}}.


The configuration space 
of the bosons \footnote{We exclusively consider bosons. Note that contact interaction between spinless fermions always vanishes, and both particle types with hard-core Coulomb interaction lead to identical densities \cite{RelationshipBosonFermion} and tunneling probabilities. Interesting differences between bosons and fermions can therefore only be expected for soft-core Coulomb interaction.} in the double-well potential is shown in Fig.~\ref{Fig1}(b). 
Region I (III) 
represents both particles in the left (right) well, and region II 
one particle in 
either well. The system
states can be 
visualized by their probability densities in configuration space [see Figs.~\ref{Fig1}(c--e)], which must be symmetric with respect to the diagonal, due to the wave function's 
bosonic symmetry $\Psi(x_1,x_2)=\Psi(x_2,x_1)$. As we show further below, important observables for the distinction between correlated and uncorrelated tunneling are the 
probabilities $P_0$, $P_1$, and $P_2$ to find zero, one, and two particles in the left well, obtained by integration of the probability density over regions III, II, and I, respectively.

For our subsequent analysis of the 
tunneling dynamics in terms of the underlying spectral structure, we consider the following protocol: 
Both particles are initially prepared in the ground state $\ket{g_\mathrm{L}}$ of the isolated left well [defined by $W_0 \rightarrow \infty$, see Fig.~\ref{Fig1}(c)], and then subjected to a potential 
quench \cite{PotentialQuench2}, i.e., an instantaneous lowering of the barrier to the finite value $W_0$. The non-equilibrium dynamics thus induced 
is obtained from the system's eigenfunctions $\ket{\phi_n}$ and eigenenergies $E_n$ (see Fig.~\ref{Fig2}), via the spectral decomposition
\begin{gather}
	\ket{\psi(t)} = \sum_n \braket{\phi_n|g_\mathrm{L}} \mathrm{e}^{-\mathrm{i} E_n t} \ket{\phi_n}.\label{Eq3}
\end{gather}
Note that the initial 
state $\ket{g_\mathrm{L}}$ decomposes 
into an entire 
range of eigenstates of the double-well system [energetically close to the initial state; see Figs.~\ref{Fig2}(a,d,g)]. To resolve the resulting dynamics, a single band approximation as performed,
e.g., in the Bose-Hubbard dimer model, would not do, and 
an 
accurate numerical 
determination of the full two-particle excitation spectrum is required. 
This is here accomplished by 
exact numerical diagonalization of the two-body Hamiltonian~\eqref{Eq1}
represented in a finite element basis \cite{BartelsFEM, SunZhouFEM} which
yields
a suitable discretization of 
configuration space and can be adapted to the different choices of $V_\mathrm{int}$
\cite{DissertationJB}. 

For all three interaction potentials, we can roughly classify the two-particle eigenstates of the problem into two types: states of type $T_{(1,1)}$, with 
probability densities 
mostly located in  region II, i.e., a high probability to find one particle in 
either well, and states of type $T_{(2,0)}$, with the probability density mainly located in regions I and III, i.e., a high probability to find both particles in the same well [see Figs.~\ref{Fig1}(c,e)].
By virtue of the Hellmann-Feynman-theorem \cite{Feynman, Hellmann}, the eigenstates'  types can be discriminated by the slopes (or ``velocities") of their energy levels $E_n$ in Figs.~\ref{Fig2}(a,d,g), given 
by their interaction energy expectation values,
\begin{equation}
	\frac{\mathrm{d}E_n}{\mathrm{d}U} = \braket{\phi_n|V_\mathrm{int}|\phi_n}.\label{Eq2}
\end{equation}

For contact interaction, particles in a type $T_{(1,1)}$ state 
do not interact, since they occupy different wells, resulting in vanishing slopes of the associated energy levels in Fig.~\ref{Fig2}(a). In contrast, particles in a type $T_{(2,0)}$ state 
probe the interaction potential, in general resulting in a non-vanishing slope, according to \eqref{Eq2}. This slope, however, vanishes asymptotically in case of infinitely strong repulsive contact interactions, $U\rightarrow \infty$.
For 
hard-
and soft-core Coulomb interaction, instead, the particles interact at any distance, such that none of the energy levels exhibits a vanishing slope for any values of $U$. 
However, particles in a type $T_{(2,0)}$ state 
are on average much closer to each other, resulting in a larger slope in Figs.~\ref{Fig2}(d,g), according to \eqref{Eq2}.

Two 
type $T_{(2,0)}$ states are of particular interest: For most interaction strengths $U$, 
two 
two-particle eigenstates 
are approximately given by the doublet states
$\ket{\pm} \propto \ket{g_\mathrm{L}} \pm \ket{g_\mathrm{R}}$, i.e., 
as balanced superpositions of the two-particle ground states $\ket{g_\mathrm{L(R)}}$ of the isolated left (right) well, and
energetically close to the initial state $\ket{g_\mathrm{R}}$ [indicated by the red lines in Figs.~\ref{Fig2}(a,d,g)].
However, since $T_{(2,0)}$ states have larger velocities than $T_{(1,1)}$ states, 
the doublet moves 
into resonance with type $T_{(1,1)}$ states, at isolated 
values of $U$,
leading to avoided crossings \cite{CohenTannoudjiQM1} [see Figs.~\ref{Fig2}(b,e,h)]. 
Type $T_{(1,1)}$ states exist either alone or in doublets of two states with approximately identical densities, one symmetric and one anti-symmetric with respect to the 
potential barrier, such that avoided crossings with the doublet $\ket{\pm}$ involve three or four eigenstates [see Figs.~\ref{Fig2}(b,e,h)]. As a result, the corresponding eigenstates are 
superpositions of $T_{(2,0)}$ 
states $\ket{\pm}$ with 
type $T_{(1,1)}$ 
state(s). 

In view of the structure and positions of these avoided crossings, Fig.~\ref{Fig2} shows qualitative differences not only between short- and long-range interactions, but 
also between hard- and soft-core Coulomb interaction. In the case of contact interaction, 
only a single avoided crossing appears,
at $U = 0$. 
It involves three states [see Fig.~\ref{Fig2}(b)], the doublet $\ket{\pm}$ 
and the energetically lowest 
type $T_{(1,1)}$ state. Given that contact and soft-core Coulomb interactions are identical for a strictly vanishing interaction strength (in contrast to hard-core Coulomb interaction, see our 
discussion below), they induce avoided crossings with identical structure at $U = 0$ [see Fig.~\ref{Fig2}(e)]. However, due to the 
long-range character of the soft-core Coulomb interaction, 
the doublet $\ket{\pm}$ continues to anti-cross with 
type $T_{(1,1)}$ states as $U$ increases, resulting in a series of avoided crossings.
Finally, 
hard-core Coulomb interaction induces further subtle differences:
Comparison of 
Figs.~\ref{Fig2}(d,f) 
and ~\ref{Fig2}(g,i) reveals the same number and structure of resonances as for the soft-core case, but the first two resonances appear clearly 
shifted to lower values of $U$. This is caused by the impenetrable core of the hard-core Coulomb interaction \footnote{This behavior is particular for one-dimensional systems and does not occur in higher-dimensional systems, since, in the latter case, the integral $\int_{|x| < \varepsilon} \frac{1}{|x|} \mathrm{d}x$ is finite. As a consequence, the two-particle wave function can take non-vanishing values at $x_1 = x_2$, with $x_i \in \mathbb{R}^d$ and $d > 1$, without infinite interaction energy.}, resulting -- in analogy to the fermionization of two bosons with strongly repulsive contact interaction \cite{DifferentInteractionsContactAnalytical} -- in higher energies of the doublet states $\ket{\pm}$ for the hard-core as compared to those for the soft-core Coulomb interaction,
in the 
neighbourhood 
of vanishing interaction \cite{DissertationJB}. In particular, this induces an anti-crossing at finite \emph{attractive} interaction, absent in the  spectral evolution for soft-core and contact interaction.

Let us now discuss the consequences of the above spectral properties for the tunneling dynamics. 
In the absence of resonances, when only the doublet states $\ket{\pm}$ (and, in particular, no type $T_{(1,1)}$ states) 
significantly contribute to the time evolution, we observe a correlated two-particle process, i.e., the particle number probabilities evolve like
\begin{gather}
	\begin{gathered}
		P_2(t) \approx [1+\cos(\omega t)]/2, \hspace{10pt} P_0(t) \approx [1-\cos(\omega t)]/2,\\
		P_1(t) \approx 0.
	\end{gathered}\label{Eq4}
\end{gather}
In particular, the dynamics are governed by a single tunneling frequency $\omega$, given by the doublet's tunneling splitting.
Examples for such two-particle processes are shown in Figs.~\ref{Fig3}(c,e).
Note that the underlying mechanism in Fig.~\ref{Fig3}(c) is \emph{not} the formation of a two-particle molecule due to an attractive interaction that tunnels like a single particle; the same tunneling process, with a qualitatively similar behavior, also occurs for repulsive interactions in Fig.~\ref{Fig3}(e). The observed behavior is better understood through \emph{energy conservation} \cite{Winkler2006}, 
which prevents the distribution of the particles over the two wells, due to the absence of a type $T_{(1,1)}$ state energetically close to the doublet $\ket{\pm}$ and to the initial state $\ket{g_\mathrm{L}}$. 

Resonances between the doublet and type $T_{(1,1)}$ states, in turn, lead to more complicated tunneling processes: In the vicinity of such resonances, there is a continuous transition from 
uncorrelated single-particle to correlated two-particle tunneling 
[see Figs.~\ref{Fig3}(a--c)]. In its purest form, such uncorrelated single-particle tunneling 
leads to 
\begin{gather}
	\begin{gathered}
		P_2(t) \approx p_1(t) p_2(t)\, , \hspace{10pt} P_0(t) \approx [1-p_1(t)][1-p_2(t)]\, ,\\
		P_1(t) \approx p_1(t)[1-p_2(t)] + [1-p_1(t)]p_2(t)\, ,
	\end{gathered}\label{Eq4}
\end{gather}
with the independent single-particle probabilities $p_1(t) = [1-\cos(\omega_1 t)]/2$ and $p_2(t) = [1-\cos(\omega_2 t)]/2$. Examples with $\omega_1 \approx \omega_2$ [induced by 
an avoided crossing with three contributing eigenstates] can be found in Figs.~\ref{Fig3}(a,f), and a case with $\omega_1 \approx 4 \omega_2$ [induced by an avoided crossing with four contributing eigenstates] is shown in Fig.~\ref{Fig3}(d).
Note that vanishing interaction of any kind, penetrable or impenetrable, implies uncorrelated single-particle tunneling induced by the resonances at $U = 0$ in Figs.~\ref{Fig2}(c,f,i).
Also note that uncorrelated single-particle tunneling typically happens on faster time scales [compare, e.g., the tunneling times $T$ in Figs.~\ref{Fig3}(a,c)], due to the 
stronger coupling [as compared to the two-particle tunneling coupling associated with the doublet  $\ket{\pm}$]
between the involved $T_{(1,1)}$ and $T_{(2,0)}$ 
states
in the vicinity of avoided crossings.


The above 
suggests a simple experimental single-particle observable to discriminate the different interaction types here studied:
The time resolved, normalized particle number 
in the left well, $N_\mathrm{L}(t) = P_2(t) + P_1(t)/2$ [note that the individual probabilities $P_n(t)$ to obtain exactly $n$ particles in the left well are two-particle observables]
can be expressed as
\begin{gather}
	N_\mathrm{L}(t) = \frac{1}{2} +  \sum_k A_k \cos(\omega_k t),\label{Eq5}
\end{gather}
with 
discrete 
frequencies $\omega_k$ contributing with amplitudes $A_k$ (both 
given in terms of $\ket{g_\mathrm{L}}$, together with the 
$\ket{\phi_n}$ and 
$E_n$
\footnote{Identifying each pair $m \neq n$ with an index $k \cong (m,n)$, the exact dependency is given by $\omega_k = |E_m - E_n|$ and $A_k = 2 c_m c_n \braket{\phi_m | N_\mathrm{L} | \phi_n}$, with $c_n \equiv \braket{\phi_n | g_\mathrm{L}}$, $\braket{\phi_m | N_\mathrm{L} | \phi_n} \equiv \braket{\phi_m | \phi_n}_\mathrm{I} + \braket{\phi_m | \phi_n}_\mathrm{II} / 2$, and the partial integrals $\braket{\cdot | \cdot}_\mathrm{I/II}$. Since $N_\mathrm{L}(0) = 1$, all amplitudes add up to $\sum_k A_k = 1/2$.}). 
In all exemplary cases shown in Fig.~\ref{Fig3}, 
only a small number of dominant frequencies $\omega_k$ and amplitudes $A_k$ contributes
significantly to 
$N_\mathrm{L}(t)$, due to the small number of eigenfunctions involved in the avoided crossings \footnote{Note that, despite being the ground state of the isolated left well, a quench from $\ket{g_\mathrm{L}}$ populates highly excited states of the total double-well system. As an example, for hard-core Coulomb interaction with $U \mathcal{L} = 0.3638$ [see Fig.~\ref{Fig3}(f)], 115 eigenfunctions $\ket{\phi_n}$ must be taken into account to ensure $\sum_n \left| \braket{\phi_n|g_\mathrm{L}} \right|^2 \geq 0.999$, despite the dynamics being well approximated by three dominant eigenstates, as can be seen in Fig.~\ref{Fig2}(i) [these three dominant eigenstates account for $95.6 \, \%$ of the total norm of $\ket{g_\mathrm{L}}$].}. These pairs $(\omega_k, A_k)$ are shown in Fig.~\ref{Fig4} as a function of 
$U$, for contact, soft-,
and hard-core Coulomb interaction.
It is apparent 
that the frequency content of $N_\mathrm{L}(t)$ faithfully reflects the presence of only one resonance (at $U=0$) for contact interaction, a sequence of resonances at $U\geq 0$ for soft-core
Coulomb, and a first resonance emerging already at attractive interactions for hard-core Coulomb interactions.

Let us conclude with the remark that the number and width of resonances will typically increase at 
stronger coupling of the potential wells (i.e, smaller values of $bW_0$), stronger long-range interactions, or larger particle numbers,
such that, ultimately, uncorrelated tunneling will prevail \cite{DissertationJB}. 
Consequently, in contrast to the here observed, primarily correlated tunneling of two moderately interacting particles, we 
expect uncorrelated or at 
best partially correlated tunneling to dominate in the 
limit of large particle numbers.

\begin{acknowledgements}
The authors thank J.-M. Rost for fruitful discussions. J. B. thanks the Studienstiftung des deutschen Volkes for support. C. D. acknowledges the Georg H. Endress foundation for support and the Freiburg Institute for Advanced Studies for a FRIAS Junior Fellowship. This work has been supported by the Baden-W\"urttemberg Stiftung gGmbH through Grant No. QT-9 NEF2D. 
\end{acknowledgements}

\bibliography{Bibliography}

\end{document}